\documentclass{article}

\usepackage{amsmath,marvosym,fullpage,bm}
\usepackage{psfrag}             
\usepackage{subfigure}          
\usepackage{graphicx}
\usepackage{wrapfig}
\usepackage{amsfonts}
\usepackage{amssymb}
\usepackage{algorithm,algorithmic}

\begin{document}

\title{A new transformative framework for data assimilation and calibration of physical ionosphere-thermosphere models}


\author{
Piyush M. Mehta\thanks{Research Associate, Department of Aerospace Engineering and Mechanics, Email: piyushmukeshmehta@gmail.com.},\
Richard Linares\thanks{Assistant Professor, Department of Aerospace Engineering and Mechanics. Email: rlinares@umn.edu.}
\\
{\normalsize\itshape
    University of Minnesota, Minneapolis, NM, USA}\\
}




%
%
%
%
%


\date{}

\maketitle

\begin{abstract}
Accurate specification and prediction of the ionosphere-thermosphere (IT) environment, driven by external forcing, is crucial to the space community. In this work, we present a new transformative framework for data assimilation and calibration of the physical IT models. The framework has two main components: (i) the development of a quasi-physical dynamic reduced order model (ROM) that uses a linear approximation of the underlying dynamics and effect of the drivers, and (ii) data assimilation and calibration of the ROM through estimation of the ROM coefficients that represent the model parameters. A reduced order surrogate for thermospheric mass density from the Thermosphere Ionosphere Electrodynamic General Circulation Model (TIE-GCM) was developed in previous work. This work concentrates on the second component of the framework - data assimilation and calibration of the TIE-GCM ROM. The new framework has two major advantages: (i) a dynamic ROM that combines the speed of empirical models for real-time capabilities with the predictive capabilities of physical models which has the potential to facilitate improved uncertainty quantification (UQ) using large ensembles, and (ii) estimation of model parameters rather than the driver(s)/input(s) which allows calibration of the model, thus avoiding degradation of model performance in the absence of continuous data. We validate the framework using accelerometer-derived density estimates from CHAMP and GOCE. The framework is a first of its kind, simple yet robust and accurate method with high potential for providing real-time operational updates to the state of the upper atmosphere in the context of drag modeling for Space Situational Awareness and Space Traffic Management.
\end{abstract}


\section{Introduction}\label{s:Intro}
The upper atmosphere, comprising the ionosphere-thermosphere (IT), is a highly dynamic environment that readily undergoes variations that can be significant under certain conditions. Accurate modeling and prediction of the IT variations, caused by space weather events (SWEs), are crucial for safeguarding the space assets that serve various communities. Ionospheric enhancements caused by SWEs can hinder telecommunications while also affecting systems on-board the assets directly through surface charging and other phenomenon. Thermospheric mass density enhancements caused by SWEs have a direct and strong impact on the drag force acting on the space assets and other objects in low Earth orbit (LEO). Existing models for the thermosphere can be highly biased or erroneous, especially for forecasts, making drag the largest source of uncertainty in our ability to accurately predict the state of the objects in LEO. With the recent increase in space traffic (\cite{MegaC}), predicting the state of the objects in LEO becomes critical for collision avoidance in the context of space situational awareness (SSA) and Space Traffic Management (STM).

Empirical models of the thermosphere (\cite{Jacchia,Bow1,Bow2,Hedin1977,Hedin1983,Hedin1987,MSIS,Barlier,Berger,Bruinsma2003,Bruinsma2012,Bruinsma2015}), developed since early in the space age using sparse measurements, adopt a climatological approach to modeling the variations of the thermosphere. These models capture the behavior in an average sense using low-order, parameterized mathematical formulations tuned to observations. A major advantage of the empirical models is that they are fast to evaluate, making them ideal for drag and SSA/STM applications. The current state of practice employed by the Joint Space Operations Center (JSpOC) under the direction of the US Air Force Space Command is an operational assimilative empirical model that makes dynamic adjustments based on recent measurements of the state of the thermosphere (\cite{Storz}) but lacks in its ability for providing accurate forecasts. 

The upper atmosphere is a large-scale nonlinear physical dynamical system with exogenous inputs, including from the Sun, which is its strongest driver. The first principles based physical models of the IT carry good potential for forecast; however, realizing that potential requires significant advances in data assimilation methods. Data assimilation is the process of fusing observational data into numerical models to reduce uncertainty in the model forecast. Data assimilation is required for physical models due to the imperfect nature of the dynamics embedded in them that allows empirical models to consistently outperform them in terms of accuracy (\cite{Shim}).

Over the last decade or two, significant advances have been made in development of data assimilation methods with IT models for neutral density and drag applications. The said methods have been successful in achieving better agreement between the measurements and model, but lack in consistently providing nowcasts and forecasts that can compete with current state of practice (\cite{Sutton_DA}) - the High Accuracy Satellite Drag Model (HASDM) (\cite{Storz}). Most methods for data assimilation achieve this either by estimating the state, the driver(s) or some combination of the two (\cite{Fuller_Rowell,Minter,Codrescu2004,Matsuo2012,Matsuo2013,Matsuo_Knipp,Morozov,Murray,Godinez,Codrescu2018,Sutton_DA}). The state can either be the parameter of interest (model output) or model parameters that relate the output to the input. Because physical models solve the discretized fluid equations over a volumetric grid, the full state can be rather large in size (over a million estimated parameters). Traditional data assimilation methods, based on the Ensemble Kalman Filter (EnKF), estimate both the input drivers as well internal state of the model because the variational time-scales in the thermosphere can cause a lag in the filter. The large state vector combined with the large number of ensembles needed to obtain statistically significant results makes the approach computationally expensive. Recent approach by \cite{Sutton_DA} uses pre-defined model variation runs in lieu of large ensembles combined with an iterative approach to prevent filter lag in estimation of the dominant drivers ($F_{10.7}$ and $K_p$) for a self-consistent calibration of the model. The approach however remains computationally expensive requiring dedicated parallel resources for real-time application, but more importantly, the estimation can result in physically unrealistic values for the driver(s) as previous methods that estimate them. In addition, the method of estimating driver(s) is not robust against a break in continuous data stream as the model forecast falls back to the original evolution of the model, and is currently an open question in the community [private conversation with Humberto Godinez]. Moreover, with the final goal of accurate uncertainty quantification for computation of collision probabilities, the current methods lie somewhere between highly computationally expensive to intractable. 

This paper demonstrates a new two-part transformative framework for data assimilation and calibration of physical IT models with the potential for providing accurate density forecasts and uncertainty quantification. The framework has two main components: (i) development of a quasi-physical reduced order model (ROM), and (ii) calibration of the ROM through data assimilation. Previous work presented development of a new method, Hermitian Space - Dynamic Mode Decomposition with control (HS-DMDc), towards achieving model order reduction for large scale dynamical IT models (\cite{Mehta_ROM}). The new method carries the same motivation and goals as previous work using Proper Orthogonal Decomposition (POD) or Empirical Orthogonal Functions (EOFs) (e.g. \cite{Mehta_POD,Matsuo2012}), but uses a dynamic systems formulation that inherently facilities prediction. The ROM provides a linearized representation of the underlying model dynamics. In this paper, we demonstrate a simple yet robust and effective approach for estimating and calibrating the state of the thermosphere using the ROM with data assimilation. The approach uses a standard Kalman Filter to estimate a reduced state that represents the model parameters rather than the driver(s), which avoids degradation of the model performance in the absence of measurement data. In addition, the ROM can provide a 24-hour forecast in a fraction of a second on a standard desktop platform. In essence, the framework combines the best of both empirical (low cost) and physical (predictive capabilities) models, and can also facilitate accurate uncertainty quantification with large ensemble simulations (subject of future work).

The paper is organized as follows: section \ref{s:ROM} describes the process of developing a quasi-physical dynamic ROM using discrete time simulation output from physical IT systems. Section \ref{s:D2C} describes the conversion of the ROM between discrete and continuous time, essential for assimilating data using a Kalman Filter in the new framework. Section \ref{s:Measurements} describes the accelerometer-derived density measurements used in this work for data assimilation. Section \ref{s:KF} gives a brief overview of the very popular Kalman Filter including the Bayesian Optimization approach used to optimize the filter in this work. Section \ref{s:Observability} answers the question of observability and section \ref{s:RD} presents and discusses the results of data assimilation. Finally, section \ref{s:Conclusions} concludes the paper.

\section{Model Order Reduction}\label{s:ROM}
The main idea behind reduced order modeling is to reduce the complexity of a physical model by reducing the state space dimension or degrees of freedom of the system. Proper Orthogonal Decomposition (POD) (\cite{Lumley}) or Empirical Orthogonal Function (EOF) is the most commonly used approach for model order reduction, however, one of its major limitation is that the formulation does not allow prediction. Several studies have attempted to overcome this limitation using surrogate or parameterized models (e.g., \cite{Mehta_POD}); however, the upper atmosphere is a physical dynamical system and should be appropriately modeled as such. Dynamic Model Decomposition (DMD), developed by \cite{DMD}, uses a dynamic systems formulation that inherently enable prediction. [Proctor et al 2014] extended DMD for application to systems with exogenous inputs. \cite{Mehta_ROM} used the Hermitian Space to extend application to large-scale systems such as the upper atmosphere; they call the method Hermitian Space - Dynamic Model Decomposition with control (HS-DMDc). They developed a ROM using simulation output from NCAR's TIE-GCM (Thermosphere Ionosphere Electrodynamics - Global Circulation Model) (\cite{TIE-GCM}) spanning over a full solar cycle (12 years). The reader is referred to \cite{Mehta_ROM} for detailed information on ROM for IT models. In this paper, we provide basic knowledge relevant to the process of data assimilation. 

Large scale physical models of the IT solve discretized fluid equations over a grid on interest. In discrete time, the evolution of a linear dynamical system can be given as
\begin{equation}\label{e:DS1}
{\bf x}_{k+1}={\bf A}_d{\bf x}_k + {\bf B}_d{\bf u}_k
\end{equation}
where ${\bf A}_d \in \mathbb{R}^{n\times n}$ is the dynamic matrix and ${\bf B}_d \in \mathbb{R}^{n\times q}$ is the input matrix in discrete time, $\textbf{x} \in \mathbb{R}^{n\times 1}$ is the full state, $\bf u \in \mathbb{R}^{q\times 1}$ is the input(s), and $k$ is the time index. HS-DMDc uses time resolved snapshots, ${\bf x}_k$, from a physical system, in this case TIE-GCM simulation output, to extract the best fit estimate for $\bf A$ and $\bf B$. In order for the derived model to be applicable across the full range of input conditions, the ROM was derived using 12 years of TIE-GCM simulations spanning a full solar cycle. Because 12 years worth of simulation output or snapshots results in a large dataset, the innovation behind HS-DMDc is to reduce the problem to the Hermitian Space (computing the inverse of a matrix $\textbf{X} \in \mathbb{R}^{n\times (m-1)}$, m being the number of snapshots, is reduced to taking an inverse of $\textbf{X}\textbf{X}^T \in \mathbb{R}^{n\times n}$). 

HS-DMDc uses snapshot matrices that are a collection of the time-resolved output from TIE-GCM to estimate the dynamic and input matrices. The three-dimensional grid outputs over time are unfolded into column vectors and stacked together. The input matrix is an assimilation of the inputs to the system. In this case, the inputs used are the solar activity proxy ($F_{10.7}$), geomagnetic proxy ($K_p$), universal time (UT) and day of the year.
\begin{equation}\label{e:HS-DMDc1}
{\bf X}_1 = \left[\begin{matrix}
| & |  &  & | \\ {\bf x}_1 & {\bf x}_2 & \cdots & {\bf x}_{m-1} \\ | & |  &  & | 
\end{matrix}\right] \quad {\bf X}_2 = \left[\begin{matrix}
| & |  &  & | \\ {\bf x}_2 & {\bf x}_3 & \cdots & {\bf x}_{m} \\ | & |  &  & | 
\end{matrix}\right] \quad {\bf \Upsilon} = \left[\begin{matrix}
| & |  &  & | \\ {\bf u}_1 & {\bf u}_2 & \cdots & {\bf u}_{m-1} \\ | & |  &  & | 
\end{matrix}\right]
\end{equation}  
The snapshot and input matrices are related by Eq.~\ref{e:DS1} such that
\begin{equation}\label{e:DS2}
{\bf X}_2={\bf A}_d{\bf X}_2 + {\bf B}_d{\bf \Upsilon}
\end{equation}
The goal now is to estimate the ${\bf A}_d$ and ${\bf B}_d$. In order to achieve this, the above equation is modified such that 
\begin{equation}\label{e:HS-DMDc2}
{\bf X}_2 = {\bf Z}\mathbf{\Psi}
\end{equation}
where ${\bf Z}$ and $\mathbf{\Psi}$ are the augmented operator and data matrices respectively. 
\begin{equation}\label{e:DMDc3}
{\bf Z} \triangleq \begin{bmatrix} {\bf A} & {\bf B}
\end{bmatrix} \quad \text{and} \quad \mathbf{\Psi} \triangleq \begin{bmatrix} {\bf X}_1 \\ 
\mathbf{\Upsilon}
\end{bmatrix}
\end{equation}
The estimate for ${\bf Z}$, and hence ${\bf A}_d$ and ${\bf B}_d$, is achieved with a Moore-Penrose pseudo-inverse of $\mathbf{\Psi}$ such that ${\bf Z} = {\bf X}_2\mathbf{\Psi}^{\dagger}$. 

Because the state size, $n$, can also be very large making storage and computation of the dynamic and input matrices intractable, a reduced state is used to model the evolution of the dynamical system. 
\begin{equation}\label{e:DS3}
{\bf z}_{k+1}={\bf A}_r{\bf z}_k + {\bf B}_r{\bf u}_k + {\bf w}_k
\end{equation}
where ${\bf A}_r \in \mathbb{R}^{r\times r}$ is the reduced dynamic matrix and ${\bf B}_r \in \mathbb{R}^{r\times q}$ is the reduced input matrix in discrete time, $\textbf{z} \in \mathbb{R}^{r\times 1}$ is the reduced state, and ${\bf w}_k$ is the process noise that accounts for the unmodeled effects. The state reduction is achieved using a similarity transform ${\bf z}_k = {\bf U}_{r}^{\dagger}{\bf x}_k = {\bf U}_{r}^{T}{\bf x}_k$, where ${\bf U}_{r}$ are the first $r$ POD modes. The steps involved in HS-DMDc are summarized below. The data assimilation process presented in this work will estimate the reduced state, $\textbf z$, that represents the coefficients of the POD modes and can be thought of as model parameters that relate the model input(s) to output(s). It can also provide insights into the model dynamics and will be explored in future work.

\begin{algorithm}
	\caption{Hermitian Space - Dynamic Mode Decomposition with control}
	
	\vspace{0.5cm}
	1. Construct the data matrices ${\bf X}_1$, ${\bf X}_2$, $\mathbf{\Upsilon}$, and $\mathbf{\Psi}$.
	
	2. Compute the pseudo-inverse of ${\bf \Psi}$ using an economy eigendecomposition (E-ED) in the Hermitian Space. The choice of $\hat{r}$ depends on several factors; in this work we set $\hat{r}$ = 20. 
	\begin{equation}
	\mathbf{\Psi}\mathbf{\Psi}^T=\hat{{\bf U}}_{\hat{r}}\hat{\mathbf{\Xi}}_{\hat{r}}\hat{{\bf U}}_{\hat{r}}^T 
	\end{equation} 
	\begin{equation}
	\Longrightarrow \mathbf{\Psi}^{\dagger}=\mathbf{\Psi}^T(\mathbf{\Psi}\mathbf{\Psi}^T)^{-1} = \mathbf{\Psi}^T(\hat{{\bf U}}_{\hat{r}}\hat{\mathbf{\Xi}}_{\hat{r}}\hat{{\bf U}}_{\hat{r}}^{\dagger})^{-1}=\mathbf{\Psi}^T\hat{{\bf U}}_{\hat{r}}\hat{\mathbf{\Xi}}_{\hat{r}}^{-1}\hat{{\bf U}}_{\hat{r}}^T
	\end{equation} 
	
	3. Perform a second E-ED in the Hermitian space to derive the POD modes (${\bf U}_{r}$) for reduced order projection. Choose the truncation value $r$ such that $\hat{r} > r$; in this work we set $r$ = 10.
	\begin{equation}
	{\bf X}_1{\bf X}_1^T = {\bf U}_{r}\mathbf{\Xi}_{r}{\bf U}_{r}^T
	\end{equation}

	4. Compute the reduced order dynamic and input matrices
	\begin{equation}
	\tilde{{\bf A}} = {\bf U}_{r}^T{\bf X}_2\mathbf{\Psi}\hat{{\bf U}}_{\hat{r}}\hat{\mathbf{\Xi}}_{\hat{r}}^{-1}\hat{{\bf U}}_{\hat{r},1}^T{\bf U}_{r}
	\end{equation}
	\begin{equation}
	\tilde{{\bf B}} = {\bf U}_{r}^T{\bf X}_2\mathbf{\Psi}\hat{{\bf U}}_{\hat{r}}\hat{\mathbf{\Xi}}_{\hat{r}}^{-1}\hat{{\bf U}}_{\hat{r},2}^T
	\end{equation}
	where $\hat{{\bf U}}_{\hat{r}}^T = [\hat{{\bf U}}_{\hat{r},1}^T \; \hat{{\bf U}}_{\hat{r},2}^T]$ with $\hat{{\bf U}}_{\hat{r},1} \in \mathbb{R}^{n \times \hat{r}}$ and $\hat{{\bf U}}_{\hat{r},2} \in \mathbb{R}^{p \times \hat{r}}$	
\end{algorithm}

\section{Discrete to Continuous time}\label{s:D2C}
The new framework uses a sequential (Kalman) filter for data assimilation that requires propagating the state to the next measurement time, which most likely will not be uniformly distributed and/or with a snapshot resolution used to derived the dynamic and input matrices for the ROM. Therefore, the dynamic and input matrices need to be first converted to continuous time and then back to time of next measurement, $t_k$. This can be achieved using the following relation (\cite{D2C})
\begin{equation}\label{e:D2C}
\begin{bmatrix} 
{\bf A}_c & {\bf B}_c\\
{\bf 0} & {\bf 0} 
\end{bmatrix} = \log \left(\begin{bmatrix} 
{\bf A}_d & {\bf B}_d\\
{\bf 0} & {\bf I} 
\end{bmatrix}\right) / {\text T}
\end{equation}
where ${\bf A}_c$ is the dynamic matrix and ${\bf B}_c$ is the input matrix in continuous time, and ${\text T}$ is the sample time (snapshot resolution when converting from discrete to continuous time and the time to next measurement, $t_k$, when converting back from continuous to discrete time). This represents another major advantage of the new framework where the time-step of model evolution can be readily adjusted. 

\section{Measurements}\label{s:Measurements}

Because the TIE-GCM ROM used for this demonstation is restricted to altitudes between 100 and 450 km for reasons discussed in \cite{Mehta_ROM}, we cannot use the pair of CHAMP (CHAllenging Minisatellite Payload) (\cite{CHAMP}) and GRACE (Gravity Recovery and Climate Experiment) (references for CHAMP and GRACE) (\cite{GRACE}) accelerometer-derived high accuracy measurements of thermospheric mass density. CHAMP and GRACE accelerometer-derived mass density estimates have been the workhorse for a lot of work that has been done in the area; therefore, we use CHAMP derived measurements in conjunction with density estimates derived from accelerometer measurements onboard the GOCE (Gravity Field and Steady-State Ocean Circulation Explorer) satellite (\cite{GOCE}). We use the state-of-the-art CHAMP density estimates from \cite{Densities} and the latest GOCE data set from \cite{GOCE_Densities}. Unlike most of the work in the area that uses some form of intercalibration process to make the assimilation and validation dataset consistent, we do not perform any intercalibration as first proposed by \cite{Mehta_AGU}. In essence, this work also provides an independent validation of the self-consistency of the two datasets, at least in reference to TIE-GCM.

Because the existing models, empirical and physical, have the largest bias/difference with accelerometer derived densities at solar minimum and geomagnetically active conditions (\cite{Densities}), we choose a representative day (of the year: 320) in November 2009 to demonstrate the new framework. This period is also chosen because GOCE was launched in March 2009 and data for both CHAMP and GOCE is available. CHAMP is in a nearly polar orbit at close to 320 km altitude on the day chosen for assimilation. GOCE is in a sun-synchronous orbit close to 250 km in altitude. We choose a day with low geomagnetic activity because of the limitation on the current version of the TIE-GCM ROM as discussed in \cite{Densities}. Demonstration for active time periods will subject of future work. Both CHAMP and GOCE derived density estimates have a time resolution of 10 seconds. As in previous work by the authors, development of the TIE-GCM ROM and assimilation of CHAMP densities in performed in the log scale (\cite{Emmert and Picone}).

\section{Kalman Filter}\label{s:KF}
The sequential (Kalman) Filter, henceforth referred to in this work as KF, has been the workhorse for state estimation and prediction using discrete-time linear systems since its inception at the beginning of the space age (\cite{KF}). The KF is a linear optimal state estimation method using statistical (Bayesian) inference based on the Bayes' theorem (\cite{Bayes}). Since the KF is one of the most commonly used tools in estimation theory, we will only provide a basic description with equations. There is a significant amount of literature available to the reader on Bayes's theorem, Bayesian inference, and KF should they be interested.

The KF has two major steps: (i) Time update, and (ii) Measurement update, as outlined in the algorithm below. The time update projects the current state and covariance estimate forward through the model to the time of next measurement. The projected state (${\bf x}_{k}^{-}$) and covariance (${\bf P}_{k}^{-}$), signified by the negative superscript, represent the \textit{apriori} knowledge about the state of the system. The process noise (${\bf Q}$) in Eq.~\ref{e:KF2} accounts for the imperfect model dynamics. The \textit{apriori} covariance, the measurement variance (${\bf R}_k$), and the observation matrix (${\bf H}_k$) are combined to compute the Kalman Gain (${\bf K}_{k}$) as given in Eq.~\ref{e:KF3}. We generate a Monte Carlo estimate for ${\bf R}_k$ in the log scale based on the uncertainties associated with the measurements ($\rho$ - mass density) by sampling the Gaussian 100 times and recomputing the variance using the samples such that ${\bf R}_k \approx \text{Var}\left[\text{log}(\rho_k + \Delta\rho_k*\text{randn}(100))\right]$. The observation matrix ${\bf H}_{k}^T \in \mathbb{R}^{r\times 1}$, which relates the measurements ($\tilde{{\bf y}}_k$) to the state (${\bf z}_k$) by mapping it onto the measurement space, is in this case a vector made up of the interpolated values at the measurement location of the first $r$ POD modes such that $\tilde{{\bf y}}_k = {\bf H}_k{\bf z}_k+{\bf v}_k$, where ${\bf v}_k$ is the measurement error. In simple terms, the Kalman Gain reflects the weights or confidence for the \textit{apriori} estimate against the measurement. The Kalman Gain is then used to update the \textit{apriori} state and covariance estimate as given in Eqs.~\ref{e:KF4} and \ref{e:KF5}. The updated state (${\bf x}_{k}^{+}$) and covariance (${\bf P}_{k}^{+}$), signified by the positive superscript, represent the \textit{posteriori} knowledge about the state of the system achieved after data assimilation. The \textit{posteriori} estimates are fed back into steps 1 and 2 until all the measurements have been processed. In the absence of measurements, the state and covariance are propagated through the model until a measurement is available.

\subsection{Bayesian Optimization}

A well-known challenge when applying the Kalman filter is the lack of knowledge on the process noise statistics. This work uses the process noise model to account for unmodeled effects that are not captured by TIE-GCM and errors induced in the model reduction process. Tuning the Kalman filter includes estimating statistics for process and measurement noise. However, we assume that the measurement noise values reported for the dataset used are accurate. Therefore, we only tune the process noise covariance. 

\begin{figure}[h]
	\centering
	\includegraphics[width=\textwidth]{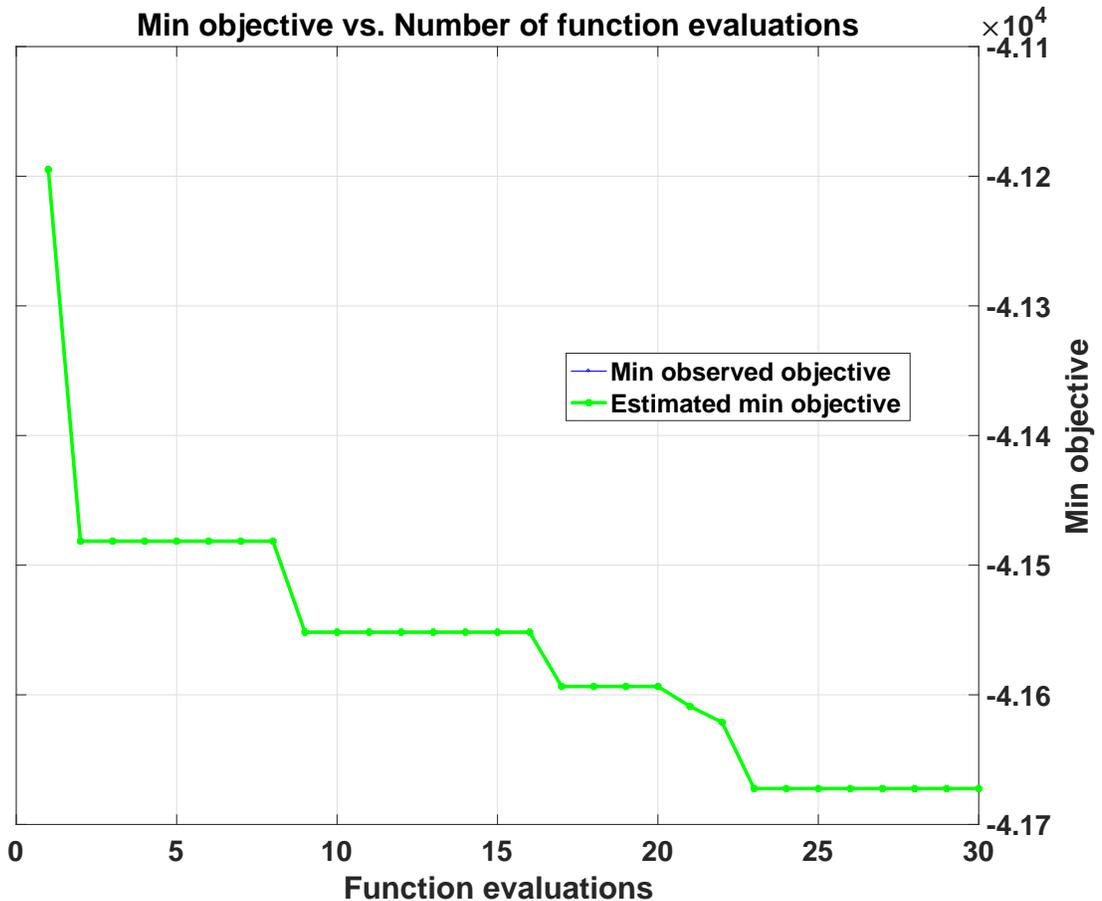}
	\caption{Cost of Bayesian optimization.}
	\label{f:Cost_Function}
\end{figure}

Bayesian Optimization (BO) is a method for blackbox optimization of stochastic cost function (\cite{BO}). We use BO because our cost is a complex function of the process noise (${\bf Q}$) and stochastic due to the measurements noise (${\bf R}$). We use the cost function for maximum likelihood estimation (MLE) given as
\begin{equation}\label{e:MLE}
\min_{\bf q} L(\tilde{\bf y}|{\bf q})=\sum_{k=1}^{N}\log\left(\det\left({\bf P}_k^{+}+{\bf H}_k{\bf R}_k{\bf H}_k^T\right)\right)+ \left(\hat{\bf y}_k-\tilde{\bf y}_k\right)^T\left({\bf P}_k^{+}+{\bf H}_k{\bf R}_k{\bf H}_k^T\right)^{-1} \left(\hat{\bf y}_k-\tilde{\bf y}_k\right)
\end{equation}
where $\hat{\bf y}_k$ is the estimated density and ${\bf Q}$ = diag(${\bf q}$), ${\bf q} \in \mathbb{R}^{r\times1}$. We use the \textit{bayesopt} function built into MATLAB for current work. Based on some initial non-optimized runs, the range of optimizable process noise ${\bf Q}$ is set at [1e-5, 5e-4]. Because we estimate the reduced state that represents the POD coefficients, we initialize the filter with an initial covariance of ${\bf P}_0$ = 10*${\bf I}_{r\times r}$. The number of iterations is set at the default of 30. Figure~\ref{f:Cost_Function} show the cost for the optimization cycle. The optimized process noise matrix ${\bf Q}_0$ is then used to initialize the KF. 

\begin{algorithm}
	\caption{Kalman Filter}
	
	\textbf{Time update}
	
	1. Project the initial state estimate forward in time 
	\begin{equation}\label{e:KF1}
	{\bf x}_{k+1}^{-} = {\bf A}{\bf x}_{k} + {\bf B}{\bf u}_{k}
	\end{equation}  
	2. Project the initial covariance estimate forward in time
	\begin{equation}\label{e:KF2}
	{\bf P}_{k+1}^{-} = {\bf A}{\bf P}_{k}{\bf A}^T + {\bf Q}
	\end{equation}
	
	\textbf{Measurement update}
	
	3. Compute the Kalman Gain
	\begin{equation}\label{e:KF3}
	{\bf K}_{k+1} = {\bf P}_{k+1}^{-} {\bf H}_{k+1}^T \left({\bf H}_{k+1}{\bf P}_{k+1}^{-}{\bf H}_{k+1}^T + {\bf R}_{k+1}\right)^{-1}
	\end{equation} 
	4. Update the State Estimate
	\begin{equation}\label{e:KF4}
	{\bf x}_{k+1}^{+} = {\bf x}_{k+1}^{-} + {\bf K}_{k+1}\left(\tilde{{\bf y}}_{k+1} -{\bf H}_{k+1}{\bf x}_{k+1}^{-}\right)
	\end{equation} 
	5. Update the Covariance Estimate
	\begin{equation}\label{e:KF5}
	{\bf P}_{k+1}^{+} = \left({\bf I} - {\bf K}_{k+1}{\bf H}_{k+1} \right) {\bf P}_{k+1}^{-}
	\end{equation} 
	
\end{algorithm}

\section{Observability Analysis}\label{s:Observability}
The new framework begs two different questions on observability: (i) Is there enough data available to derive/extract the POD modes for model order reduction, and (ii) Can the reduced state, $\bf z$, be estimated using discrete point measurements of density along an orbit. The first question is easy to answer because unlike some previous works where the POD modes or EOFs are derived from discrete measurements of density along an orbit (e.g., \cite{EOF}), we use model simulation output that provides complete global spatial coverage to derive the POD modes. In addition, simulation output over a full solar cycle is used which ensures sufficient temporal coverage.

\begin{figure}[h]
	\centering
	\includegraphics[width=\textwidth]{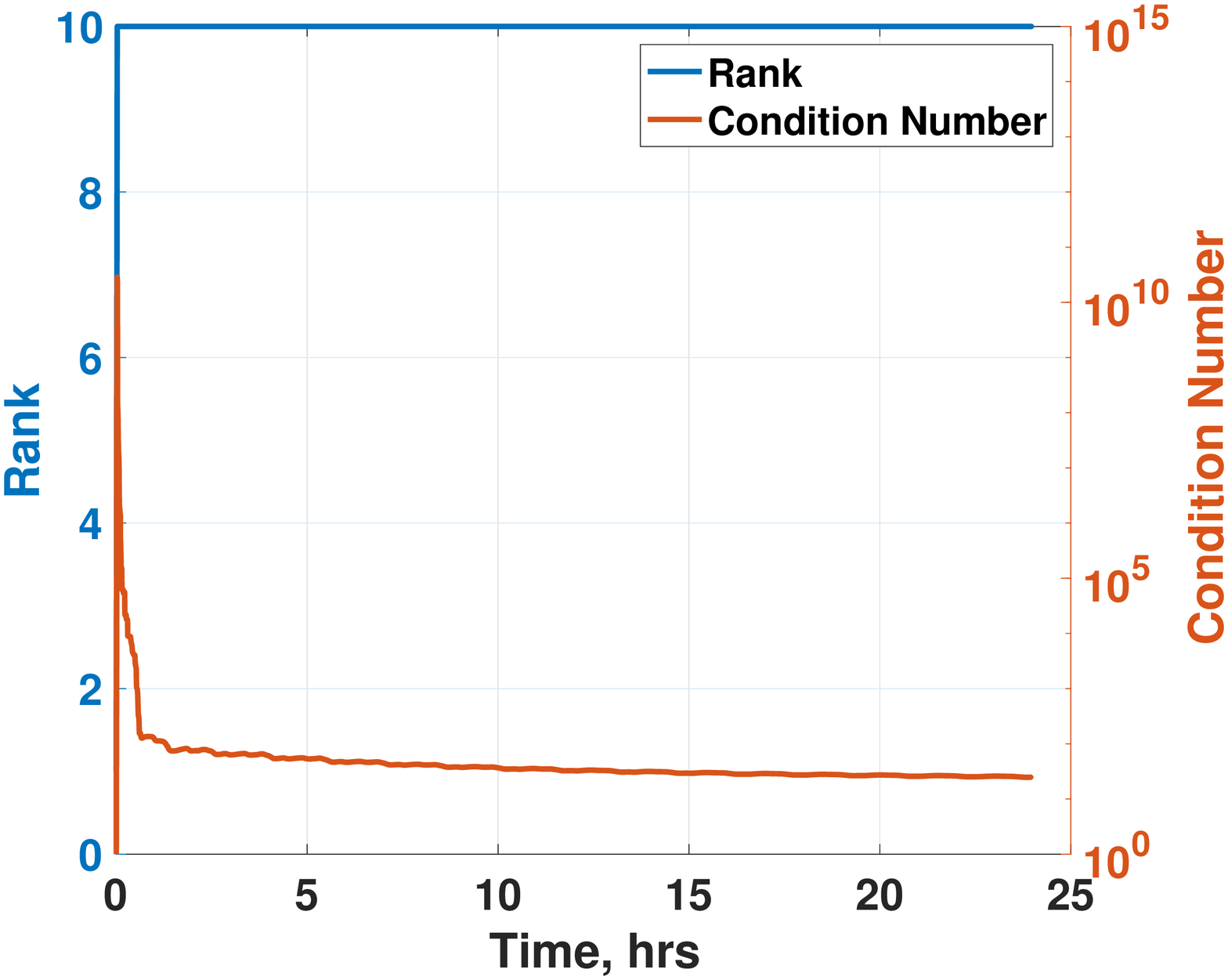}
	\caption{Rank and Condition number of the Observability Matrix when assimilating CHAMP measurements with TIE-GCM ROM on day 320 of year 2009.}
	\label{f:RANK_CN}
\end{figure}

For linear time-invariant systems, such as the ROM formulation in Eq.~\ref{e:DS1}, answer to the second question can be provided using the observability matrix
\begin{equation}\label{e:OM}
\pmb{\mathcal{O}}_{k} = {\bf H}_{r,k}{\textbf A}_{r}^{k} \quad \quad \Longrightarrow \quad \quad \pmb{\mathcal{O}} = 
\begin{bmatrix} 
{\bf H}_0 \\ {\bf H}_{1}{\textbf A} \\ {\bf H}_{2}{\textbf A}^{2} \\ \vdots
\end{bmatrix} 
\end{equation}
The matrix $\pmb{\mathcal{O}}$ is populated at each time for the location at which the observation is made. The state $\bf z$ is considered to be observable when the observability matrix is full rank or rank($\pmb{\mathcal{O}}$) = $r$. In addition, the observability matrix can artificially become full rank because of noise. Therefore, we also compute the condition number of $\pmb{\mathcal{O}}$ to confirm that this is not the case. A relatively large condition number suggests that the observability matrix is not full rank because of noise. Figure~\ref{f:RANK_CN} shows the rank and condition number for the demonstration case presented in this work assimilating a full days' worth of measurements along the CHAMP orbit for $r$ = 10. The figure shows that the observability matrix becomes full rank very quickly. The condition number rises to very large values initially (large values correspond to ill-conditioned problem) but also falls back to a well-conditioned state quickly after full rank is achieved. This ill-conditioned period corresponds to the transition time of the Kalman Filter.

\section{Results and Discussion}\label{s:RD}
We demonstrate the new framework by assimilating CHAMP accelerometer-derived densities on day 320 of 2009. We assimilate CHAMP measurements and validate with an independent set of measurements from GOCE. The discrete-time reduced order dynamic and input matrices are converted to continuous time using the relation in Eq.~\ref{e:D2C}.  Since, both the ingested and validation dataset have a time resolution of 10 seconds, we convert the dynamic and input matrices back to discrete time for a time-step of 10 seconds. We perform assimilation using $r$ = 10 modes, based on the results in \cite{Mehta_ROM}. 

Another major advantage of the new framework is that the ROM can be initialized using output from any model. Therefore, TIE-GCM simulations are not required in case of an interruption in operations. The model can easily be re-initialized using output from empirical models. For this demonstration, we choose to initialize the ROM using model output from the Naval Research Laboratory's MSIS model (\cite{MSIS}). The global density (full state) from the MSIS model is used to compute the reduced state as ${\bf z}_0 = {\bf U}{\bf x}_0$.

\begin{figure}[h]
	\centering
	\includegraphics[width=1.1\textwidth]{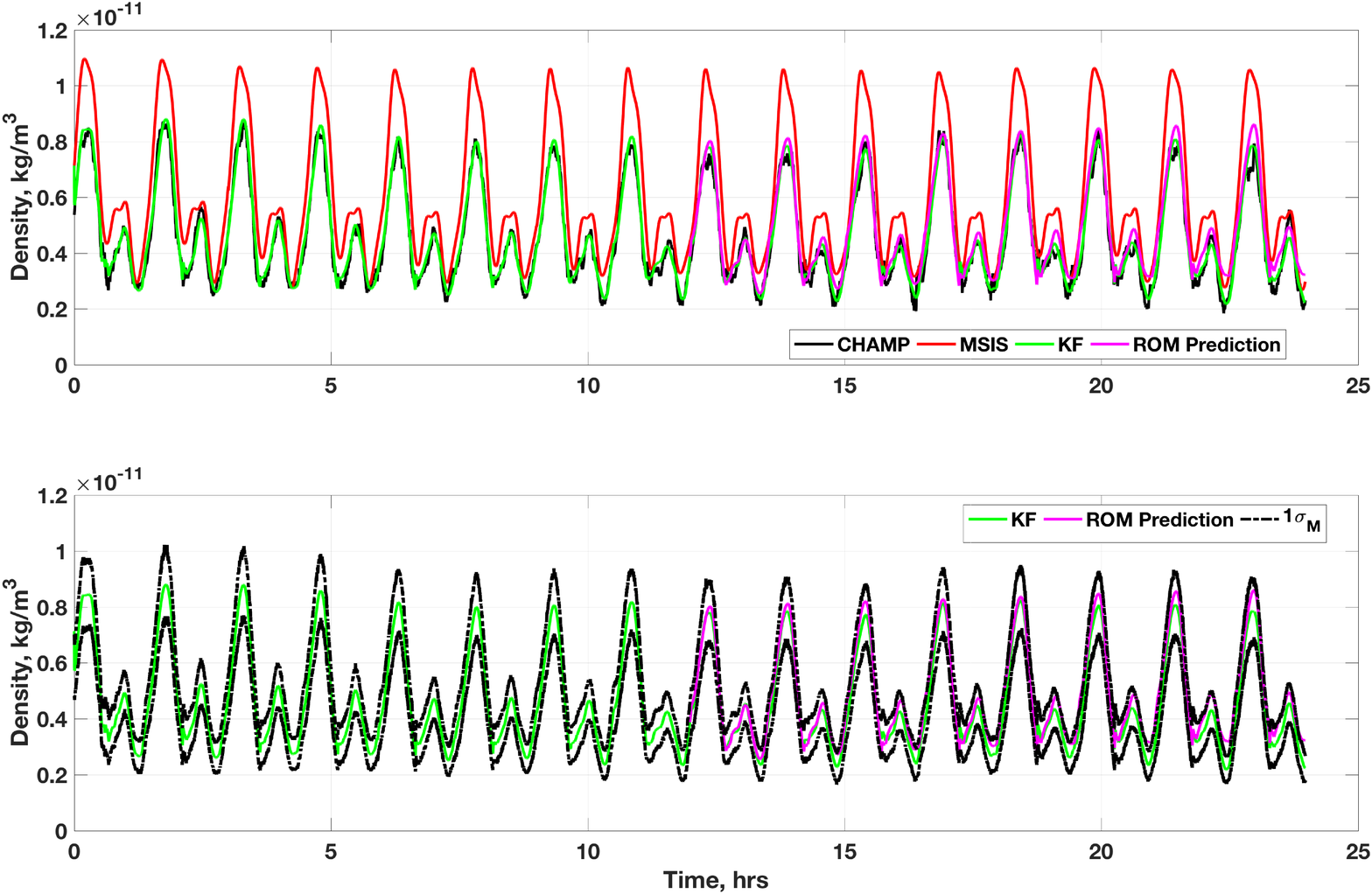}
	\caption{(top) CHAMP assimilated ROM densities on day 320 for year 2009. Magenta: prediction with ROM after 12 hours of data assimilation. (bottom) KF estimated and ROM predicted densities with KF estimated 1$\sigma$ uncertainties along CHAMP orbit.}
	\label{f:Density_CHAMP}
\end{figure}

\begin{figure}[h]
	\centering
	\includegraphics[width=1.1\textwidth]{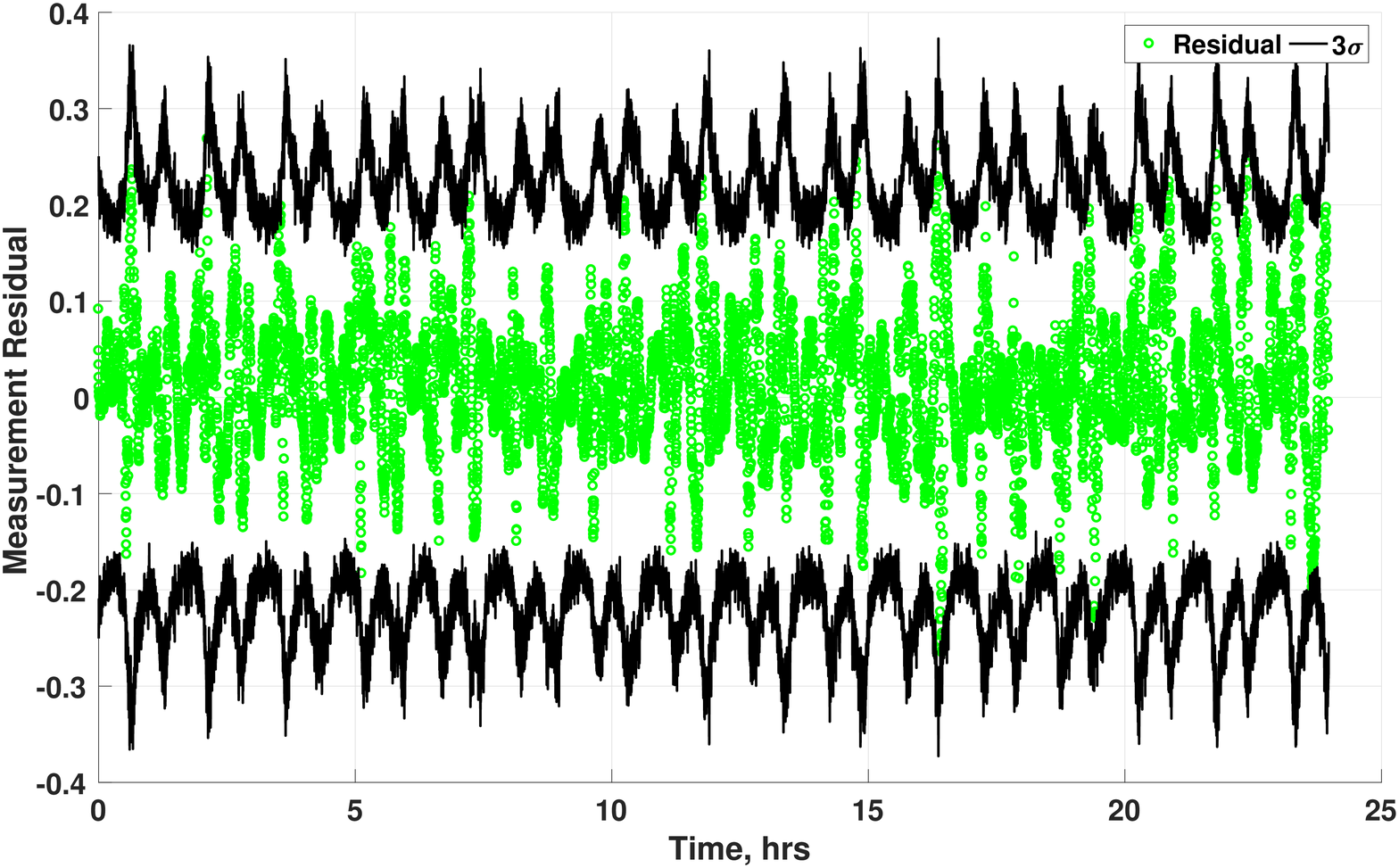}
	\caption{CHAMP measurement residuals as a result of data assimilation on day 320 of year 2009.}
	\label{f:CHAMP_Residuals}
\end{figure}

Figure \ref{f:Density_CHAMP} shows effectiveness of data assimilation process achieved with the new framework. The top panel shows the CHAMP measurements in black, MSIS in red for comparison, and the assimilated densities in green. In order to show that the model is calibrated and performs well even without available data, we run an additional ensemble that sees data assimilation for the first 12 hours but is then allowed to evolve under the dynamics captured by the ROM. This prediction is shown in magenta. As can be seen, the approach not only tracks the measurements very well, but possesses very good potential for providing accurate forecasts. The approach can correct for the absolute scale biases, which represents the major component of the errors due to drag. In this case, because the ROM is initialized with MSIS, the filter calibrates to lower values of density since MSIS is known to overpredict during times of low solar activity. Had the filter been initialized with simulation output from TIE-GCM (physical model), the data assimilation process would have resulted in a scaling up of the density. The bottom panel shows the estimated state (green) and the predicted state (magenta) with 1$\sigma$ covariance bounds estimated as part of the data assimilation. The uncertainty bounds are computed by projecting the state covariance of Eq.~\ref{e:KF5} onto the measurement space
\begin{equation}\label{e:Den_UQ}
{\bf P}_{k}^{yy} = \left({\bf H}_k{\bf P}_{k}{\bf H}_{k}^T + {\bf R}_k\right)^{0.5}
\end{equation} 
As seen, both the estimated and predicted states lie within the estimated 1$\sigma$ uncertainties bounds. Figure \ref{f:CHAMP_Residuals} shows the measurement residual for the ingested dataset. The residuals are well behaved and bounded by the 3$\sigma$ estimated uncertainty bounds. MSIS root mean square error with respect to the CHAMP measurements is 1.6e-12 while post assimilation the rms error is 2.71e-13.

\begin{figure}[h]
	\centering
	\includegraphics[width=1.1\textwidth]{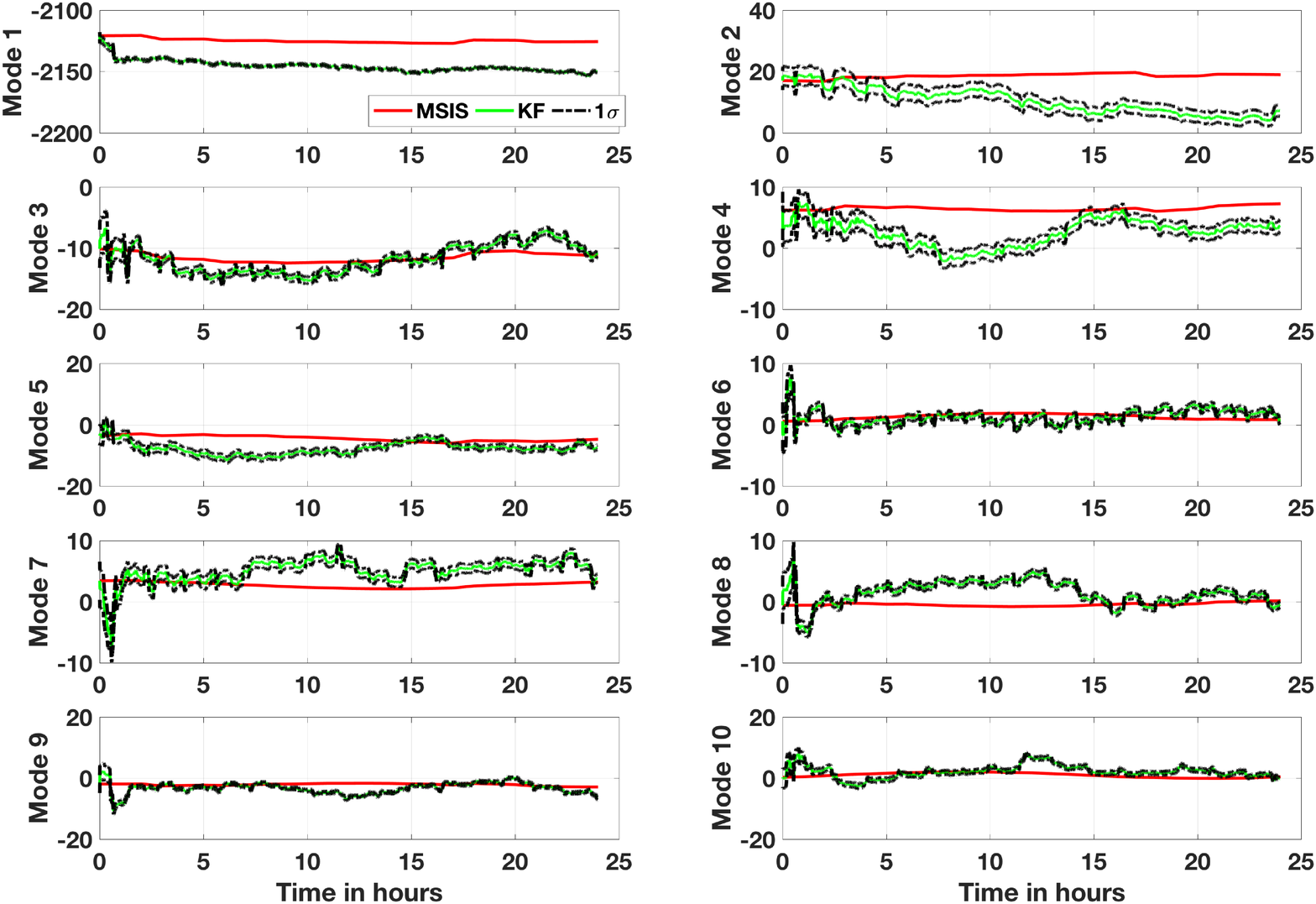}
	\caption{Red: POD coefficients for the first 10 modes computed by projecting MSIS simulation output onto the POD modes ${\bf U}_r$ on day 320 of year 2009. Green: KF estimated reduced order state ${\bf z}$ with data assimilation. Black: KF estimated 1$\sigma$ uncertainty for the reduced order state.}
	\label{f:Estimated_State}
\end{figure}

Figure \ref{f:Estimated_State} shows the estimated reduced order state and uncertainty with data assimilation. As previously discussed, the reduced state represents the POD coefficients for the first $r$ modes used for order reduction. We also show the POD coefficients for MSIS obtained by projecting the simulation output for the day onto the POD modes. The first two modes correspond to absolute scale correction (scaling with solar activity), while others that seem to oscillate about the MSIS state suggest the average nature of the empirical models.

\begin{figure}[h]
	\centering
	\includegraphics[width=1.1\textwidth]{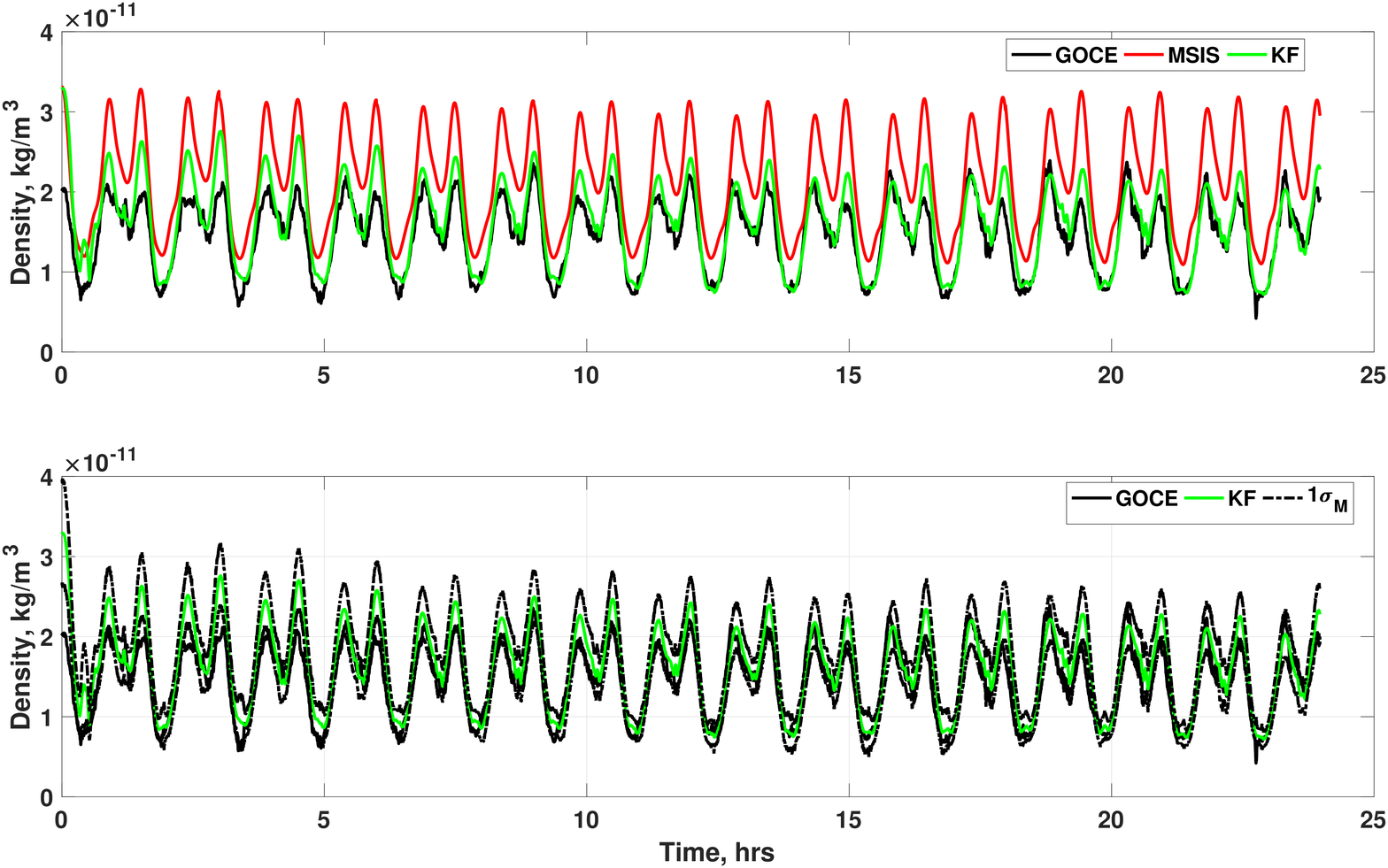}
	\caption{(top) Validation of CHAMP data assimilation using independent measurements along GOCE orbit on day 320 of year 2009. (bottom) KF estimated density and 1$\sigma$ uncertainty along GOCE orbit.}
	\label{f:Density_GOCE}
\end{figure}

Figure~\ref{f:Density_GOCE} shows the validation of the data assimilation using an independent dataset of GOCE accelerometer-derived mass density. Again, MSIS density along GOCE orbit is also shown for reference. The validation confirms that the reduced state, which provides global calibration, can be estimated using discrete measurements along a single orbit. The ROM is tuned with CHAMP densities, with the corrected state accurately predicting the density along GOCE orbit. Results show that the approach can self-consistently calibrate the model while preserving the underlying dynamics. The results also inadvertently validate the two data sets, CHAMP and GOCE, against each other in reference to the TIE-GCM model. It reflects the importance of using the state-of-the-art density datasets of \cite{Densities} for CHAMP and GRACE. The bottom panel shows that the GOCE measurements lie within the estimated 1$\sigma$ bounds about the estimated density. The authors want to point out that the rms of uncertainties provided with the GOCE dataset was on the order of 1\%, which is too small even for the best drag coefficient and gas surface interaction models (\cite{Mehta_JSR}). Therefore, we multiply the uncertainty by a factor of 5 to bring it to more reasonable values. MSIS root mean square error with respect to the GOCE measurements is 5.62e-12 while post assimilation the rms error is 2.56e-12.

\begin{figure}[h]
	\centering
	\includegraphics[width=1.1\textwidth]{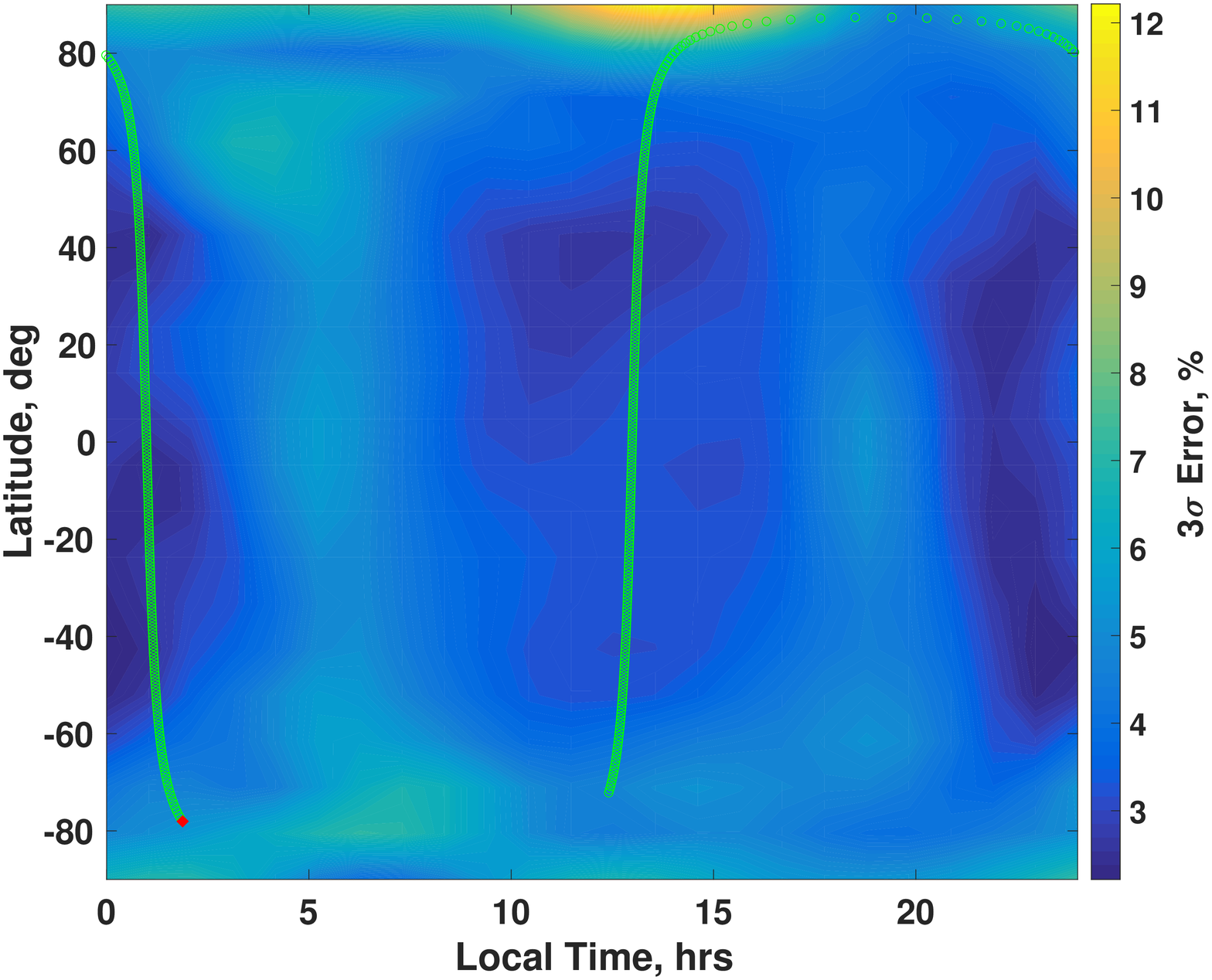}
	\caption{3$\sigma$ error prediction at mean CHAMP altitude after assimilation of all CHAMP data on day 320 of year 2009. Red: Satellite location at the current time. Green: back propagated CHAMP orbit from current location.}
	\label{f:Global_Error}
\end{figure}

\begin{figure}[h]
	\centering
	\includegraphics[width=1.1\textwidth]{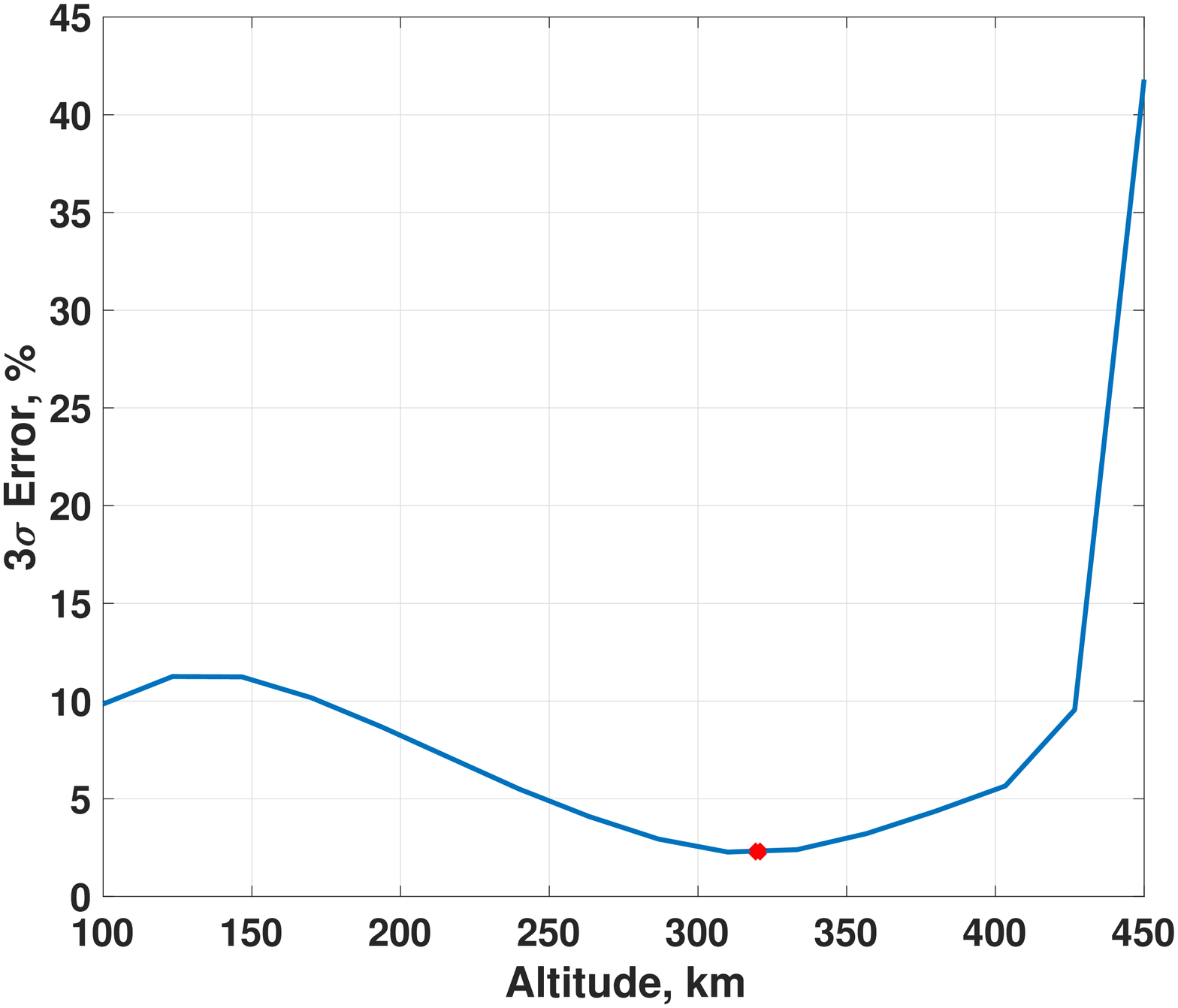}
	\caption{3$\sigma$ error prediction as a function of altitude after assimilation of all CHAMP data on day 320 of year 2009. Red: Satellite location at the current time.}
	\label{f:Altitude_Error}
\end{figure}

Figures~\ref{f:Global_Error} and \ref{f:Altitude_Error} shows the global estimated covariance as a projection away from the location of measurements. The green curve in Figure~\ref{f:Global_Error} represents CHAMP orbit path with the red point corresponding to the current location. The global uncertainty estimate is generated after all of the data has been assimilated. The global field is generated using Eq. \ref{e:Den_UQ}, but without the measurement error ${\bf R}$ and ${\bf H}$ computed over the grid at the mean altitude of CHAMP. As seen, the uncertainty is reduced in the vicinity of the satellite path where the measurements are assimilated. The uncertainty is the largest at the pole because of the singularity constraint. Figure~\ref{f:Altitude_Error} shows that the error is minimum at the altitude of CHAMP but increases moving away from the assimilated path. Even though ingesting data along an orbit path can provide global estimates using the new framework, the global errors can be further reduced with improved spatial and temporal coverage of measurements.

\begin{figure}[h]
	\centering
	\includegraphics[width=1.1\textwidth]{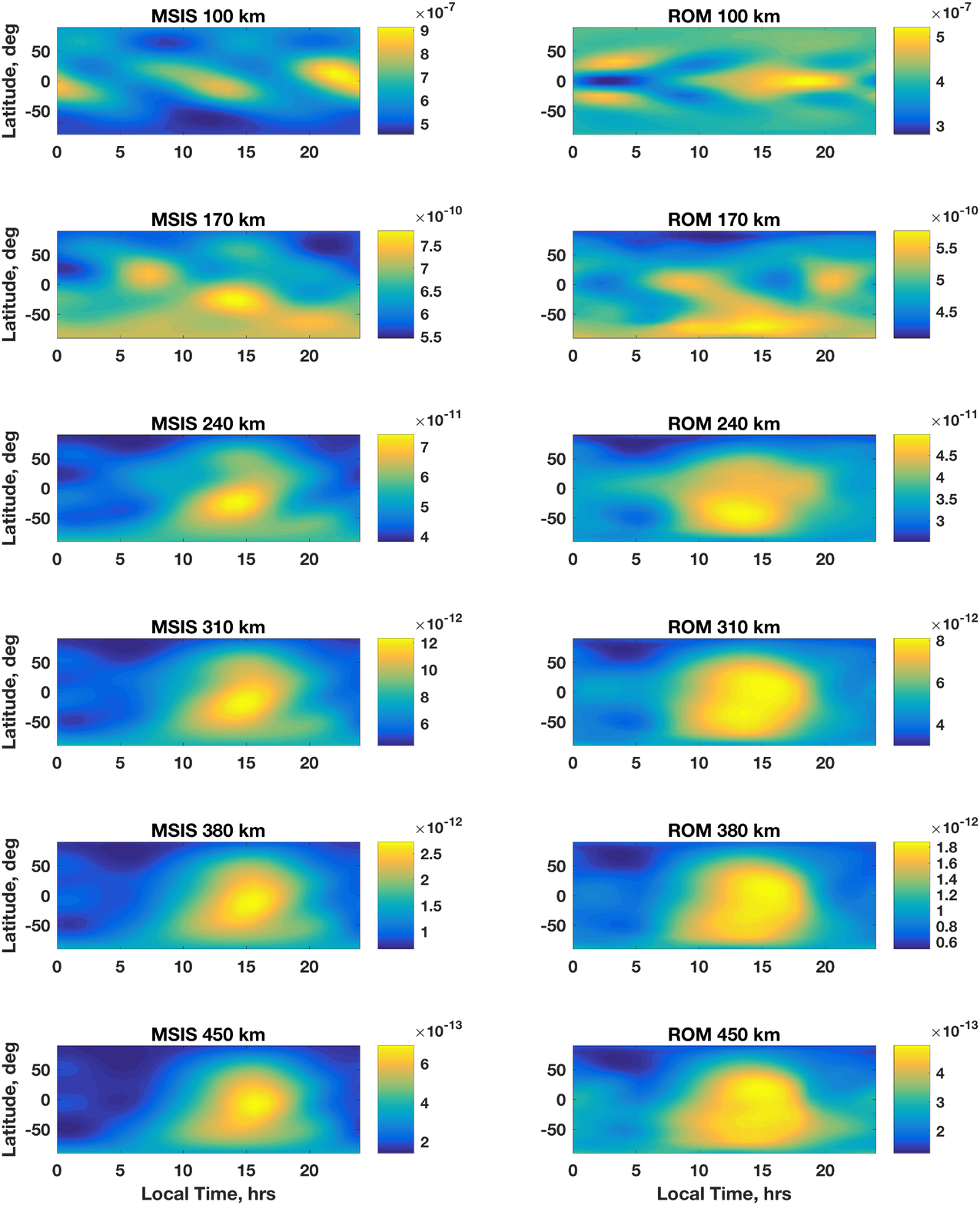}
	\caption{Comparison of mass density profiles at different altitudes from MSIS (left column) again assimilated ROM mass density profiles at the same altitudes (right column) on day 320 of year 2009.}
	\label{f:Altitude_Compare}
\end{figure}

Figure~\ref{f:Altitude_Compare} shows the comparison of MSIS profiles against the ROM assimilated profiles at a series of altitudes. As seen, except for the profile at 100 km, the MSIS and assimilated ROM profiles show similar distributions. The difference at 100 km is due to the lower boundary effects. The absolute scale of the assimilated densities suggests that MSIS overpredicts the mass density across all altitudes during periods of low solar activity. While this is not a new revelation, the new framework effectively calibrates existing physical models by adjusting the absolute scale, which is a major driver of orbit prediction errors.

\section{Conclusions}\label{s:Conclusions}
This paper has demonstrated a new, transformative framework for data assimilation and calibration of physical ionosphere-thermosphere models. A robust yet simple and effective approach for data assimilation has thus far eluded the community. The new framework has two major components: (i) Model order reduction for a quasi-physical linear representation of the dynamics, and (ii) data assimilation calibration of the reduced order model (ROM) using a Kalman Filter. Development of a ROM was discussed in a previous paper. This paper demonstrates the second component.

The new framework combines the best of both empirical and physical models. The ROM reduces the cost of model evaluation to the level of empirical models while inherently providing forecast/predictive capabilities. Unlike large scale physical models, the ROM formulation allows rapid modifications in the time-step of model evaluation or simulation with a negligible increase in the computational cost. This allows the model to be easily projected to the time of next measurement. The ROM formulation also allows large ensemble runs of the models for improved characterization and quantification of forecast uncertainty, a crucial requirement for accurate computation of collision probabilities. 

Data assimilation using the new framework also carries several advantages. The data assimilation process estimates a reduced state that represents model parameters. Therefore, the data assimilation self-consistently brings the model to agreement with measurements without modifying the model dynamics. Also, estimating model parameters rather than the input(s)/driver(s) allows the model to be calibrated. The calibration avoids degradation of model performance in the absence of measurement data.

The data assimilation cycle and prediction for a full day takes only a fraction of a section using minimal computational resources. Therefore, the framework can be readily incorporated into operations. The demonstration here is limited to non-storm time conditions. Future work will include development of ROMs that capture the nonlinear dynamics and demonstration of the data assimilation and prediction during storm time. The CHAMP and GOCE data used in this work are not considered operational datasets, therefore future work will demonstrate the effectiveness of the framework with operational datasets such as Two-Line Element (TLE) sets for well-behaved objects in LEO or other objects using recent advances for improved characterization of object parameters that affect drag (eg. \cite{Mehta_JSR_2018}) and computation of physical drag coefficients for complex objects (\cite{RSM}).

%

\section{acknowledgments}
The authors wish to acknowledge support of this work by the Air Force's Office of Scientific Research under Contract Number FA9550-18-1-0149 issued by Erik Blasch. The authors wish to thank Humberto Godinez for insightful discussion on existing methods for data assimilation. The authors also wish to thank the anonymous reviewers for their helpful comments. No data was generated as part of this work. The CHAMP densities used in this work can be downloaded at http://tinyurl.com/densitysets. The GOCE densities data is provided by the European Space Agency and can be downloaded at https://earth.esa.int/web/guest/missions/esa-operational-missions/goce/air-density-and-crosswind-data. The TIE-GCM ROM used in this work can be downloaded at the University of Minnesota Digital Conservancy: http://hdl.handle.net/11299/194705.


\begin{thebibliography}{}
	
\bibitem[{\textit{Barlier et al.,}}(1978)]{Barlier} 
Barlier, F., C. Berger, J. L. Falin, G. Kockarts, and G. Thuillier (1978), A thermospheric model based on satellite drag data, Annales de Geophysique, 34, 9-24.

\bibitem[{\textit{Bayes}}(1763)]{Bayes} 
Bayes, T., (1763). "An essay towards solving a Problem in the Doctrine of Chances." Bayes's essay as published in the Philosophical Transactions of the Royal Society of London, Vol. 53, p. 370, on Google Books.

\bibitem[{\textit{Berger et al.,}}(1998)]{Berger} 
Berger, C., R. Biancale, F. Barlier, and M. Ill (1998), Improvement of the empirical thermospheric model DTM: DTM94 - a comparative review of various temporal variations and prospects in space geodesy applications, Journal of Geodesy, 72, 161-178, doi:10.1007/s001900050158.

\bibitem[{\textit{Bruinsma et al.,}}(2003)]{Bruinsma2003} 
Bruinsma, S., G. Thuillier, and F. Barlier (2003), The DTM-2000 empirical thermosphere model with new data assimilation and constraints at lower boundary: accuracy and properties, Journal of Atmospheric and Solar-Terrestrial Physics, 65, 1053-1070, doi: 10.1016/S1364-6826(03)00137-8.

\bibitem[{\textit{Bruinsma et al.,}}(2012)]{Bruinsma2012} 
Bruinsma, S. L., N. Sanchez-Ortiz, E. Olmedo, and N. Guijarro (2012). Evaluation of the DTM-2009 thermosphere model for benchmarking purposes, Journal of Space Weather and Space Climate, 2 (27), A04, doi:10.1051/swsc/2012005.

\bibitem[{\textit{Bruinsma et al.,}}(2015)]{Bruinsma2015} 
Bruinsma, S. (2015), The DTM-2013 thermosphere model, Journal of Space Weather and Space Climate, 5 (27), A1, doi:10.1051/swsc/2015001.

\bibitem[{\textit{Bowman et al.,}}(2008a)]{Bow1} 
Bowman, B. R., W. Kent Tobiska, F. A. Marcos, and C. Valladares (2008a), The JB2006 empirical thermospheric density model, Journal of Atmospheric and Solar-Terrestrial Physics, 70, 774-793, doi:10.1016/j.jastp.2007.10.002.

\bibitem[{\textit{Bowman et al.,}}(2008b)]{Bow2} 	
Bowman, B. R., W. K. Tobiska, F. A. Marcos, C. Y. Huang, C. S. Lin, and W. J. Burke (2008b), A new empirical thermospheric density model JB2008 using new solar and geomagnetic indices, in AIAA/AAS Astrodynamics Specialist Conference and Exhibit, Honolulu, Hawaii.

\bibitem[{\textit{Codrescu et al.,}}(2004)]{Codrescu2004} 	
Codrescu, M. V., T. J. Fuller-Rowell, and C. F. Minter (2004), An ensemble-type Kalman Fillter for neutral thermospheric composition during geomagnetic storms, Space Weather, 2, S11002, doi:10.1029/2004SW000088.

\bibitem[{\textit{Codrescu et al.,}}(2018)]{Codrescu2018} 	
Codrescu, S. M., M. V. Codrescu, and M. Fedrizzi (2018). An ensemble kalmanFillter for the thermosphere-ionosphere, Space Weather, 16, 57-68, doi:10.1002/2017SW001752, 2017SW001752.

\bibitem[{\textit{DeCarlo}}(1989)]{D2C} 	
DeCarlo, R. A., (1989). Linear systems: a state variable approach with numerical implementation Prentice-Hall, Inc. Upper Saddle River, NJ, USA, ISBN:0-13-536814-6.

\bibitem[{\textit{Doornbos et al.,}}(2014)]{GOCE_Densities} 
Doornbos, E., S. Bruinsma, B. Fritsche, G. Koppenwallner, P. Visser, J. Van Den IJssel, and J. de Teixeira de Encarnacao (2014), GOCE+ Theme 3: Air density and wind retrieval using GOCE data, Final Report, TU Delft, Netherlands.

\bibitem[{\textit{Drinkwater et al.,}}(2003)]{GOCE} 
Drinkwater, M. R., Floberghagen, R., Haagmans, R., Muzi, D., and Popescu, A., 2003. GOCE: ESA's first Earth explorer core mission. In Beutler, G., Drinkwater, M. R., Rummel, R., and von Steiger, R. (eds.), Earth Gravity Field from Space - From Sensors to Earth Sciences. Dordrecht: Kluwer. Space Sciences Series of ISSI, Vol. 17, pp. 419-432. ISBN 1-4020-1408-2.

\bibitem[{\textit{Emmert and Picone}}(2010)]{Emmert and Picone} 
Emmert, J. T., and Picone, J. M., (2010). Climatology of globally averaged thermospheric mass density. \textit{J. Geophys. Res.}, 115, A09326, http://dx.doi.org/10.1029/2010JA015298.

\bibitem[{\textit{Fuller-Rowell et al.,}}(2004)]{Fuller_Rowell} 	
Fuller-Rowell, T. J., C. F. Minter, and M. V. Codrescu (2004), Data assimilation for neutral thermospheric species during geomagnetic storms, Radio Science, 39, RS1S03, doi:10.1029/2002RS002835.

\bibitem[{\textit{Godinez et al.,}}(2015)]{Godinez} 	
Godinez, H. C., E. Lawrence, D. Higdon, A. Ridley, J. Koller, and A. Klimenko (2015), Specification of the ionosphere-thermosphere using the ensemble kalman Filter, in Dynamic Data-Driven Environmental Systems Science, edited by S. Ravela and A. Sandu, pp. 274-283, Springer International Publishing, doi:10.1007/978-3-319-25138-7 25.

\bibitem[{\textit{Hedin et al.,}}(1977)]{Hedin1977} 
Hedin, A. E., C. A. Reber, G. P. Newton, N. W. Spencer, J. E. Salah, J. V. Evans, D. C. Kayser, D. Alcayde, P. Bauer, and L. Cogger (1977), A global thermospheric model based on mass spectrometer and incoherent scatter data MSIS. I-N2 density and temperature, J. Geophys. Res., 82, 2139-2147, doi:10.1029/JA082i016p02139.

\bibitem[{\textit{Hedin}}(1983)]{Hedin1983} 	
Hedin, A. E. (1983), A revised thermospheric model based on mass spectrometer and incoherent scatter data MSIS-83, J. Geophys. Res., 88, 10, 170-10 ,188, doi: 10.1029/JA088iA12p10170.

\bibitem[{\textit{Hedin}}(1987)]{Hedin1987} 
Hedin, A. E. (1987), MSIS-86 thermospheric model, J. Geophys. Res., 92, 4649-4662, doi:10.1029/JA092iA05p04649.

\bibitem[{\textit{Jacchia}}(1970)]{Jacchia} 
Jacchia, L. G. (1970), New Static Models of the Thermosphere and Exosphere with Empirical Temperature Proles, SAO Special Report, 313.

\bibitem[{\textit{Kalman}}(1960)]{KF} 
Kalman, R. E. (1960). "A New Approach to Linear Filtering and Prediction Problems". Journal of Basic Engineering. 82: 35. doi:10.1115/1.3662552.

\bibitem[{\textit{Lumley}}(1967)]{Lumley} 
Lumley, J. L., (1967). The structure of inhomogeneous turbulent flows, \textit{Proceedings of the International Colloquium on the Fine Scale Structure of the Atmosphere and its Influence on Radio Wave Propagation}, edited by A. M. Yaglam and V. I. Tatarsky, Doklady Akademii Nauk SSSR, Moscow, Nauka.

\bibitem[{\textit{Matsuo and Forbes}}(2010)]{EOF} 
Matsuo, T., and Forbes, J. M., (2010). Principal Modes of Thermospheric Density Variability: Empirical Orthogonal Function Analysis of CHAMP 2001-2008 data. \textit{J. Geophys. Res.}, 115, A07309, doi:10.1029/2009JA015109. 

\bibitem[{\textit{Matsuo et al.,}}(2012)]{Matsuo2012} 	
Matsuo, T., M. Fedrizzi, T. J. Fuller-Rowell, and M. V. Codrescu (2012), Data assimilation of thermospheric mass density, Space Weather, 10, 05002, doi:10.1029/2012SW000773.

\bibitem[{\textit{Matsuo and Knipp et al.,}}(2013)]{Matsuo_Knipp} 	
Matsuo, T., and D. J. Knipp (2013), Thermospheric mass density specification: Synthesis of observations and models, AFRL Tech. Rep., DTIC ADA592729.

\bibitem[{\textit{Matsuo et al.,}}(2013)]{Matsuo2013} 	
Matsuo, T., I.-T. Lee, and J. L. Anderson (2013), Thermospheric mass density specification using an ensemble Kalman Filter, Journal of Geophysical Research (Space Physics), 118, 1339-1350, doi:10.1002/jgra.50162.

\bibitem[{\textit{Mehta et. al.}}(2014a)]{Mehta_JSR} 
Mehta, P. M., Walker, A. C., McLaughlin, C. A., and Koller, J., (2014a). "Comparing Physical Drag Coefficients Computed Using Different Gas-Surface Interaction Models", \textit{Journal of Spacecraft and Rockets}, Vol. 51, No. 3 (2014), pp. 873-883. 
https://doi.org/10.2514/1.A32566.

\bibitem[{\textit{Mehta et. al.}}(2014b)]{RSM} 
Mehta, P. M., Walker, A. C., Lawrence, E., Linares, R. L., Higdon, D., and Koller, J., (2014b). "Modeling satellite drag coefficients with response surfaces", \textit{Advances in Space Research}, Vol. 54, No. 8, pp. 1590-1607, https://doi.org/10.1016/j.asr.2014.06.033.

\bibitem[{\textit{Mehta et. al.}}(2017a)]{Densities} 
Mehta, P.~M., Walker, A., Sutton, E., and, Godinez, H., (2017a). New density estimates derived using accelerometers on-board the CHAMP and GRACE satellites. \textit{Space Weather}, in press, http://dx.doi.org/10.1002/2016SW001562.

\bibitem[{\textit{Mehta and Linares}}(2017b)]{Mehta_POD} 
Mehta, P. M., and R. Linares (2017b). A methodology for reduced order modeling and calibration of the upper atmosphere, \textit{Space Weather}, 15, doi:10.1002/2017SW001642.

\bibitem[{\textit{Mehta et al.,}}(2017c)]{Mehta_AGU}
Mehta, P. M., Linares, R. L., and Sutton, E. K., (2017c). Data-driven Inference and Investigation of Thermosphere Dynamics and Variations, \textit{American Geophysical Union}, Fall Meeting SA31A-2559, 11-15 Dec, New Orleans, LA.

\bibitem[{\textit{Mehta et al.,}}(2018a)]{Mehta_JSR_2018} 
Mehta, P. M., Linares, R. L., and Walker, A. C., (2018a). Photometric Data from Non-Resolved Objects for Improved Drag and Re-entry Prediction, \textit{Journal of Spacecraft and Rockets}, in press.

\bibitem[{\textit{Mehta et al.,}}(2018b)]{Mehta_ROM} 
Mehta P. M., Linares, R. L., Sutton E., K., (2018b). A quasi-physical dynamic reduced order model for thermospheric mass density via Hermitian Space Dynamic Model Decomposition, Space Weather, accepted.

\bibitem[{\textit{Minter et al.,}}(2004)]{Minter} 	
Minter, C. F., T. J. Fuller-Rowell, and M. V. Codrescu (2004), Estimating the state of the thermospheric composition using Kalman Filtering, Space Weather, 2, S04002, doi:10.1029/2003SW000006.

\bibitem[{\textit{Mockus}}(1975)]{BO} 
Mockus J. (1975). On bayesian methods for seeking the extremum. In: Marchuk G.I. (eds) Optimization Techniques IFIP Technical Conference Novosibirsk, July 1-7, 1974. Optimization Techniques 1974. Lecture Notes in Computer Science, vol 27. Springer, Berlin, Heidelberg. doi: https://doi.org/10.1007/3-540-07165-2(55).

\bibitem[{\textit{Morozov et al.,}}(2013)]{Morozov} 	
Morozov, A. V., A. J. Ridley, D. S. Bernstein, N. Collins, T. J. Hoar, and J. L. Anderson (2013), Data assimilation and driver estimation for the Global Ionosphere- Thermosphere Model using the Ensemble Adjustment Kalman Filter, Journal of Atmospheric and Solar-Terrestrial Physics, 104, 126-136, doi:10.1016/j.jastp.2013.08.016.

\bibitem[{\textit{Murray et al.,}}(2015)]{Murray} 		
Murray, S. A., E. M. Henley, D. R. Jackson, and S. L. Bruinsma (2015), Assessing the performance of thermospheric modeling with data assimilation throughout solar cycles 23 and 24, Space Weather, 13, 220-232, doi:10.1002/2015SW001163.

\bibitem[{\textit{Picone et. al.}}(2002)]{MSIS} 
Picone, J.~M., Hedin, A.~E., Drob, D.~P., (2002). NRLMSISE-00 empirical model of the atmosphere: statistical comparisons and scientific issues. \textit{J. Geophys. Res.} 107 (A12), http://dx.doi.org/10.1029/2002JA009430.

\bibitem[{\textit{Picone et. al.}}(2005)]{TLE} 
Picone, J. M., Emmert, J. T., and Lean, J. L., (2005). Thermospheric densities derived from spacecraft orbits: accurate processing of two-line element sets. \textit{J. Geophys. Res.}, 110, A03301, http://dx.doi.org/10.1029/2004JA010585.

\bibitem[{\textit{Qian et. al.,}}(2014)]{TIE-GCM}
Qian, L., A. G. Burns, B. A. Emery, B. Foster, G. Lu, A. Maute, A. D. Richmond, R. G. Roble, S. C. Solomon, and W. Wangm, (2014). The NCAR TIE-GCM: A community model of the coupled thermosphere/ionosphere system, in Modeling the Ionosphere-Thermosphere System, \textit{AGU Geophysical Monograph Series}, 2014.

\bibitem[{\textit{Radtke et al.,}}(2017)]{MegaC} 
Radtke, J., Kebschull, C., and Stoll, E., (2017). Interactions of the space debris environment with mega constellations - Using the example of the OneWeb constellation. \textit{Acta Astronautica}, 131, pp. 55-68, https://doi.org/10.1016/j.actaastro.2016.11.021.

\bibitem[{\textit{Reigber et al.,}}(2002)]{CHAMP} 
Reigber, C., Luhr, H., and Schwintzer, P., (2002). CHAMP mission status. \textit{Advances in Space Research}, 30, 129-134, doi:10.1016/S0273-1177(02)00276-4.

\bibitem[{\textit{Schmid}}(2010)]{DMD} 
Schmid, P., J., (2010). Dynamic mode decomposition of numerical and experimental data. \textit{Journal of Fluid Mechanics}, 656, pp. 5-28, doi:10.1017/S0022112010001217.

\bibitem[{\textit{Shim et. al.}}(2014)]{Shim} 
Shim, J. S., M. Kuznetsova, L. Rasttter, D. Bilitza, M. Butala, M. Codrescu, B. A. Emery, B. Foster, T. J. Fuller-Rowell, J. Huba, A. J. Mannucci, X. Pi, A. Ridley, L. Scherliess, R. W. Schunk, J. J. Sojka, P. Stephens, D. C. Thompson, D. Weimer, L. Zhu, D. Anderson, J. L. Chau, and E. Sutton (2014), Systematic evaluation of ionosphere/thermosphere (IT) models, in Modeling the IonosphereThermosphere System, pp. 145-160, John Wiley \& Sons, Ltd, doi:10.1002/9781118704417.ch13.

\bibitem[{\textit{Storz et. al.}}(2005)]{Storz} 
Storz, M. F., Bowman, B. R., Branson, J. I., Casali, S. J., and Tobiska, W. K.., (2005). High Accuracy Satellite Drag Model (HASDM). \textit{Advances in Space Research}, 36 (12), 2497-2505, http://dx.doi.org/10.1016/j.asr.2004.02.020.

\bibitem[{\textit{Sutton}}(2018)]{Sutton_DA} 
Sutton E., K., (2018). A new method of physics-based data assimilation for
the quiet and disturbed thermosphere, Space Weather, accepted.

\bibitem[{\textit{Tapley et al.,}}(20)]{GRACE} 
Tapley, B. D., S. Bettadpur, M.Watkins, and C. Reigber (2004), The gravity recovery and climate experiment: Mission overview and early results, Geophysical Research Letters, 31, L09607, doi:10.1029/2004GL019920.
	
\end{thebibliography}
\end{document}